\documentclass{article}

\usepackage{arxiv}

\usepackage{natbib}
\usepackage[utf8]{inputenc} 
\usepackage[T1]{fontenc}    
\usepackage{hyperref}       
\usepackage{url}            
\usepackage{booktabs}       
\usepackage{amsfonts}       
\usepackage{nicefrac}       
\usepackage{microtype}      
\usepackage{lipsum}
\usepackage{graphicx}
\graphicspath{ {./images/} }
\usepackage{paracol}

\title{Is manual software optimization a thing of the past?}

\author{
  Pavlin G. Poli\v{c}ar\thanks{Corresponding author: \texttt{pavlin.policar@fri.uni-lj.si}}
  \And
  Martin \v{S}pendl
  \And
  Toma\v{z} Ho\v{c}evar
  \AND
  \normalfont Faculty of Computer and Information Science\\
  University of Ljubljana\\
  Večna pot 113, 1000 Ljubljana, Slovenia
}
\date{}

\begin{document}

\maketitle

\begin{abstract}
\mbox{}\textbf{Context:} Scientific software is increasingly required to process larger datasets while maintaining acceptable execution times. Software optimization traditionally requires substantial expertise in programming, algorithms, and numerical methods. Recent advances in large language models (LLMs) offer the possibility of automating much of this process. \\
\textbf{Objectives:} We investigate whether LLM-based agents can autonomously achieve substantial performance improvements in scientific software, including mature implementations that have already been extensively optimized by human developers. \\
\textbf{Methods:} We tasked an LLM-based agent with optimizing software for three computational problems: t-SNE, single-sample gene set enrichment analysis (ssGSEA), and graphlet counting. Humans defined the scope, correctness criteria, and a verification mechanism, after which the agent worked autonomously, in some cases for several hours. Code maintainers reviewed each resulting implementation and verified its correctness.\\
\textbf{Results:} The optimized implementations were faster in all tested configurations, by up to two orders of magnitude over the fastest existing tools. The improvements included low-level code optimizations, mathematical reformulations, and an entirely new algorithm for graphlet counting.\\
\textbf{Conclusion:} Software optimization can increasingly be delegated to autonomous agents, with the human role shifting from implementing optimizations to deciding which software to optimize, defining objectives, providing verification mechanisms, and ensuring the correctness of the final software. For well-scoped, verifiable problems, we argue that manual software optimization may be a thing of the past.
\end{abstract}


\section{Introduction}

Scientific research increasingly relies on software for data analysis, simulation, and other computational tasks~\citep{EzerWhitaker2019}.
As datasets grow, the efficiency of this software increasingly determines which analyses are feasible. Yet optimizing non-trivial scientific software requires expertise in software engineering, numerical computing, and algorithm design, as well as considerable development effort.

Recent advances in large language models (LLMs) have enabled agents that can generate, modify, test, and benchmark software. Performance can be improved at different levels, from low-level code optimizations, such as better memory access patterns, vectorization, or parallelization, to reformulating the computation mathematically or replacing the algorithm altogether. Whether LLM agents can autonomously find improvements beyond the low-level ones, even for well-scoped problems with verifiable results, remains unclear, as existing evaluations report mostly surface-level gains~\citep{press2025algotune}.

We investigate this question on three computational problems: t-SNE, single-sample gene set enrichment analysis (ssGSEA), and graphlet counting.
In each case, humans defined the scope, correctness criteria, and a verification mechanism, after which an LLM agent worked without further intervention. The resulting implementations were faster in every case study, by up to two orders of magnitude over the fastest existing tools, with improvements spanning all three levels.

\section{Related Work}

Early LLM-based optimizers wrapped the model in fixed pipelines. SysLLMatic~\citep{peng2026sysllmatic} profiles a program to locate hotspots and rewrites them using a curated catalog of optimization patterns, while PIE~\citep{shypula2024learning} fine-tunes models on pairs of slower and faster C++ programs. Evolutionary systems that pair LLMs with automated evaluators have gone further, discovering new algorithms~\citep{novikov2025alphaevolve}. More recently, general-purpose coding agents have been applied to real scientific software, with the largest gains coming from rewriting code in faster languages or for GPUs~\citep{li2026scientific}.

Nevertheless, the effectiveness of general-purpose LLM-based agents remains uncertain.
On AlgoTune~\citep{press2025algotune}, current models achieved only modest speedups over mature numerical libraries and failed to discover algorithmic improvements.
On benchmarks built from real optimization commits in scientific Python libraries, agents reach less than a quarter of the expert speedup~\citep{ma2025swefficiency}, and other evaluations report negligible gains~\citep{sarikayak2026evaluating}.
In contrast, we show that a general-purpose agent can reduce the expected asymptotic cost of a computation, outperform an expert-designed algorithm, and speed up mature, hand-optimized software several-fold in place.

\section{Case studies}

To demonstrate that agentic software optimization extends beyond low-level code changes, we tasked Claude Opus~5, running in Claude Code, with either improving an existing implementation, producing a new, faster implementation of an existing algorithm, or producing a novel algorithm altogether for three real-world computational problems.
Each implementation was reviewed by domain experts, including the original developers of openTSNE and Orca, and verified to be correct.

\begin{figure*}[t]
    \centering
    \includegraphics{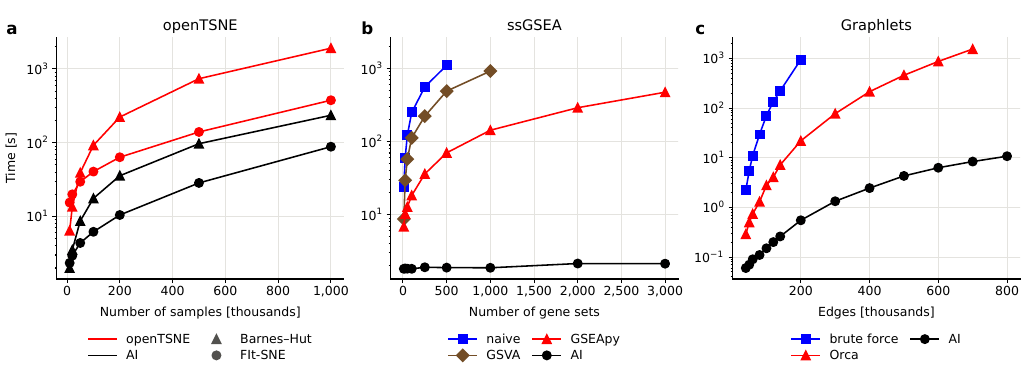}
    \caption{Runtime of existing and agent-optimized implementations on a consumer-grade M1 MacBook Pro. \textbf{(a)} t-SNE optimization phase (8 threads) on subsamples of a single-cell data set. \textbf{(b)} ssGSEA implementations on 1,000 samples with increasing gene sets. \textbf{(c)} Graphlet counting implementations on random (Erdős--Rényi) graphs with 10k nodes and an increasing number of edges.}
    \label{fig:benchmarks}
\end{figure*}

\subsection{openTSNE}

$t$-distributed stochastic neighbor embedding ($t$-SNE)~\citep{Maaten2008} is widely used for visualizing high-dimensional data.
Because its exact computation scales quadratically with the number of data points, practical applications rely on approximations.
The two most widely used approximations are Barnes--Hut, which reduces the asymptotic complexity to $\mathcal{O}(N\log N)$, and FIt-SNE, which further reduces it to $\mathcal{O}(N)$.

openTSNE~\citep{Policar2024} implements both approximations and is a mature, highly optimized CPU implementation. We tasked an agent to optimize its optimization phase, which typically accounts for 80--90\% of its total runtime, while preserving the public API, parameter defaults, optimization procedure, and accuracy of both approximations. It ran the project's 174 unit tests after each modification and retained only changes that improved the measured runtime.

After an overnight run, the optimized implementation was faster in every tested configuration, with speedups of 4.4--7.7$\times$ at 500,000 points. Single-threaded runtime decreased from 62 to 8 minutes for Barnes--Hut and from 9 to 2 minutes for FIt-SNE. On eight threads, the corresponding runtimes decreased from 12 to under 2 minutes and from 140 to 28 seconds (Figure~\ref{fig:benchmarks}.a).

The agent implemented and evaluated approximately 60 optimization ideas, retaining 34. The improvements included low-level optimizations such as specialized code paths, improved memory layout, and additional parallelism. The agent also derived a mathematical simplification of one stage of the FIt-SNE approximation, reducing the runtime of that stage by about a quarter. None changed the overall asymptotic complexity. This demonstrates that substantial gains can still be obtained through constant-factor optimization of mature software implementations.

\subsection{Single-sample GSEA}

Gene set enrichment analysis (GSEA) tests whether predefined groups of genes, such as biological pathways, are concentrated at the top or bottom of a list of genes ranked by differential expression. Single-sample GSEA (ssGSEA)~\citep{barbie2009ssgsea} applies this idea to individual samples, scoring every gene set in every sample and thereby converting a gene expression matrix into a matrix of gene set activities. ssGSEA has become a standard step in gene expression analysis, routinely used to estimate cell-type composition and to stratify samples into molecular subtypes.

Existing implementations focus on accelerating the scoring of individual sample--gene set pairs. GSEApy implements a fast kernel but still requires a pass over all ranked genes of the sample. GSVA uses a closed-form expression that needs only the ranks of the set's own genes, reducing asymptotic complexity in theory, but not realizing this gain in its implementation.
We tasked an agent with producing a new implementation of ssGSEA, requiring its scores to exactly match those of existing libraries.

Building on GSVA's closed form, the agent restructured the computation as a sparse matrix multiplication and eliminated inefficiencies, realizing this asymptotic gain.
For 2,000 gene sets scored across 1,000 samples, the runtime decreased from 5 minutes with GSEApy, the fastest existing implementation, to 2.2 seconds, a speedup of 136$\times$, with the gap widening with increasing numbers of gene sets (Figure~\ref{fig:benchmarks}.b). This makes it practical to score entire gene set collections rather than a preselected subset of pathways, removing a trade-off that analysts routinely make for computational reasons.

\subsection{Graphlet counting}

Graphlets~\citep{Przulj2004} are small connected graphs that occur as induced subgraphs of a network and can characterize its local topology. Counting the different roles that nodes can occupy within graphlets, known as orbits, provides a finer-grained description of node topology. Graphlet and orbit counts have been widely used in network analysis, including for node classification, anomaly detection, and recommendation.
Because the number of subgraphs around a node grows rapidly with its degree, brute-force enumeration quickly becomes infeasible for large or dense networks.

Orca~\citep{Hocevar2014}, a widely used orbit-counting method, avoids brute-force enumeration by deriving the counts of larger graphlets from those of smaller ones through combinatorial relations.
For graphlets with up to five nodes, this reduces the expected time complexity from $\mathcal{O}(nd^4)$ to $\mathcal{O}(nd^3)$, where $n$ is the number of nodes and $d$ the maximum node degree.
We tasked an agent with producing a new implementation with improved scaling of orbit counting, requiring its counts to match those of Orca and brute-force enumeration exactly.

After running autonomously for several hours, the agent produced an entirely new algorithm that was faster than Orca on every tested network.
On a human protein--protein interaction network from BioSNAP,\footnote{Stanford Biomedical Network Dataset Collection: \url{https://snap.stanford.edu/biodata/datasets/10000/10000-PP-Pathways.html}} with approximately 22k nodes and 340k edges, the new implementation counted all orbits in 13 seconds, compared with 35 minutes for Orca, a speedup of about 160$\times$. Brute-force enumeration would require an estimated 3.5 days, over 20,000 times longer than the new implementation.
On random graphs with 10k nodes, the gap widened with network density, and the runtime grew roughly quartically with the number of edges for brute-force enumeration, cubically for Orca, and quadratically for the agent's implementation (Figure~\ref{fig:benchmarks}.c).

\section{Conclusion}

Our three case studies demonstrate that autonomous LLM-based agents can substantially speed up scientific software, starting from mature and naive implementations alike. The improvements were not limited to surface-level optimizations; agents applied low-level code optimizations, derived mathematical reformulations, and developed an entirely new algorithm for graphlet counting. The resulting implementations were faster in every case study, in some cases by orders of magnitude.

These findings suggest that software optimization can increasingly be delegated to autonomous agents. Improvements of this kind have typically required specialists in numerical computing or algorithm design, whom most scientific projects cannot afford. Autonomous agents put them within reach of individual researchers and small research groups. The human role is therefore shifting from implementing optimizations to deciding which software to optimize, defining scope and objectives, providing verification mechanisms, and ensuring that the final software is correct.
Although problems with open-ended objectives or hard-to-verify correctness remain untested, we argue that for well-scoped, verifiable problems, manual software optimization may indeed be a thing of the past.

\section*{Declaration of competing interest}

The authors declare no competing interests.

\section*{Reproducibility}

We provide reusable optimization prompts, the new implementations, and replication materials at \url{https://github.com/pavlin-policar/llm-software-optimization}.

\section*{Funding}

This work was supported by the research programme P2-0209 of the Slovenian Research and Innovation Agency and the Project Grant L7-70273, and Young Research Grant 57111.

\bibliographystyle{abbrv}

\bibliography{main}

\end{document}